\documentclass[twocolumn,prl,10pt,showpacs,aps,footinbib]{revtex4-1}
\usepackage{graphicx,color,multirow,bm,braket,amsmath}
\usepackage[final]{pdfpages}
\newcommand{\Frohl}{Fr\"ohlich\ }
\newcommand{\TiO}{TiO$_2$\ }

\begin{document}

\title{\Frohl electron-phonon vertex from first principles}

\author{Carla Verdi}
\author{Feliciano Giustino}
\email{feliciano.giustino@materials.ox.ac.uk}
\affiliation{Department of Materials, University of Oxford, Parks Road, Oxford OX1 3PH, United Kingdom}


\begin{abstract}
We develop a method for calculating the electron-phonon vertex in polar semiconductors and insulators
from first principles. The present formalism generalizes the Fr\"ohlich vertex to the case of anisotropic 
materials and multiple phonon branches, and can be used either as a post-processing correction to standard 
electron-phonon calculations, or in conjunction with {\it ab initio} interpolation based on maximally 
localized Wannier functions. We demonstrate this formalism by investigating the electron-phonon interactions 
in anatase TiO$_2$, and show that the polar vertex significantly reduces the electron lifetimes 
and enhances the anisotropy of the coupling. The present work enables {\it ab initio} calculations of carrier 
mobilities, lifetimes, mass enhancement, and pairing in polar materials.
\end{abstract}

\maketitle

The electron-phonon interaction (EPI) is a cornerstone of condensed matter physics, and plays important 
roles in a diverse array of phenomena. Recent years have witnessed a surge of interest in 
{\it ab initio} calculations of EPIs, leading to new techniques and many innovative applications 
in the case of metals and non-polar semiconductors~\cite{Giustino2008nature, Mauri2012, Mauri2007, Park2008, 
Giustino2007Bdoped, Giustino2010diam, Marini2011, Gonze2014diam, Patrick2013, Lueders2005, Needs2014, Louie2014PRL}. 
In contrast to this fast-paced progress, in the case of polar semiconductors and insulators the study 
of EPIs from first principles has not gone very far, owing to the prohibitive computational costs of 
EPI calculations for polar materials. For example, a fully {\it ab initio} calculation of the carrier 
mobility of a polar semiconductor has not been performed yet, while such calculations have recently 
been reported for non-polar semiconductors such as silicon \cite{Restrepo2009} and graphene \cite{Bonini2014}. 
Given the fast-growing technological importance of polar semiconductors, from light-emitting devices 
to transparent electronics, solar cells and photocatalysts \cite{Pimpuktar2009, Snaith2014, Fujishima1972}, 
developing accurate and efficient computational methods for studying EPIs in these systems 
is of primary importance.

At variance with metals and non-polar semiconductors, in polar materials two or more atoms in the 
unit cell carry nonzero Born effective charge tensors \cite{BornHuang}. 
As a consequence, the fluctuations of the ionic positions corresponding to longitudinal optical (LO) phonons at long 
wavelength generate macroscopic electric fields which can couple strongly to electrons and holes, 
leading to the so-called \Frohl interaction \cite{Frohlich1954}. Up to now this interaction has not been taken 
into account in {\it ab initio} calculations of EPIs; the two key obstacles towards a description 
of \Frohl coupling from first principles are (i) the \Frohl coupling was designed to describe simple
isotropic systems with one LO phonon, and (ii) the electron-phonon vertex diverges for ${\bf q}\rightarrow 0$, 
where ${\bf q}$ is the phonon wavevector. The first obstacle relates to the fundamental question on 
how to define the \Frohl coupling in the most general way. The second obstacle renders first-principles 
calculations extremely demanding, since a correct description of the singularity requires a very 
fine sampling of the Brillouin zone. 

In this work we address the challenges (i) and (ii) above by developing a general formalism
for first-principles calculations of the \Frohl vertex. Our strategy consists in separating 
the short-range and the long-range contributions to the electron-phonon matrix elements, and identifying
the \Frohl coupling with the long-range component. We translate our formalism into a powerful 
computational scheme, whereby the short-range component is calculated using state-of-the-art Wannier-Fourier 
electron-phonon interpolation \cite{Giustino2007}, and the singular coupling is calculated
using the Born effective charges and the high-frequency dielectric permittivity tensor. As a 
first demonstration of this approach we calculate carrier lifetimes in anatase TiO$_2$.

The \Frohl model~\cite{Frohlich1954} describes the interaction of an electron in a parabolic band 
with a dispersionless LO phonon of frequency $\omega_{\rm LO}$. 
The electron is in an isotropic dielectric medium with static and high-frequency permittivities 
$\epsilon_0$ and $\epsilon_\infty$, respectively. In this model the electron-phonon coupling matrix 
element takes the form:
  \begin{equation} \label{M}
  g_{\bf q}= \frac{i}{|{\bf q}|}\left[\frac{e^2}{4\pi\varepsilon_0}   
  \frac{4\pi}{N\Omega}\frac{\hbar\omega_{\rm LO}}{2}
  \left(\frac{1}{\epsilon_\infty}-\frac{1}{\epsilon_0}\right)\right]^{\!\frac{1}{2}}\!,
  \end{equation}
where ${\bf q}$ is the phonon wavevector, $\Omega$ the unit cell volume, $N$ the number of unit
cells in the Born-von K\'arm\'an supercell, and $e$, $\varepsilon_0$, and $\hbar$ are the electron 
charge, vacuum permittivity, and reduced Planck constant, respectively. Equation~(\ref{M}) shows 
that the \Frohl coupling $g_{\bf q}$ diverges at long wavelengths, ${\bf q}\rightarrow 0$. This
singularity poses a challenge to {\it ab initio} calculations of EPIs in polar materials.

In general the vertex describing electron-one~phonon interactions can be expressed via the
coupling matrix element $g_{mn\nu}(\mathbf k,\mathbf q)= \bra{\psi_{m\mathbf k+\mathbf q}}
\Delta_{\mathbf q\nu} V\ket{\psi_{n\mathbf k}}$. This quantity has the meaning of probability 
amplitude for the scattering between the initial electronic state $\ket{\psi_{n\mathbf k}}$ 
and the final state $\ket{\psi_{m\mathbf k+\mathbf q}}$ via the perturbation $\Delta_{\mathbf q\nu}V$ 
due to a phonon  with crystal momentum 
${\bf q}$, branch $\nu$ and frequency $\omega_{\mathbf q\nu}$. 
The matrix elements $g_{mn\nu}(\mathbf k,\mathbf q)$ can be calculated 
starting from density functional perturbation theory \cite{Baroni2001}, 
and have been employed to investigate many properties involving EPIs, for example the electron 
velocity renormalization \cite{Giustino2007graphene} and lifetimes \cite{Giustino2009graphene, Echenique2003}, 
phonon softening \cite{Mauri2007} and lifetimes \cite{Liao2015ph, Mauri2006}, 
phonon-assisted absorption \cite{Kioupakis2010, Noffsinger2012}, critical temperature in 
conventional superconductors \cite{Choi2002, Lueders2005, Margine2014}, 
and resistivity \cite{Restrepo2009, Bonini2014}. The key ingredient of all these calculations is the evaluation of 
$g_{mn\nu}(\mathbf k,\mathbf q)$ on extremely dense Brillouin zone grids, which is 
computationally prohibitive. This difficulty has been overcome with the development 
of an {\it ab initio} interpolation strategy for the electron-phonon vertex \cite{Giustino2007, Giustino2010} 
based on maximally-localized Wannier functions \cite{Marzari1997, *Souza2001}. 
The approach of Ref.~\onlinecite{Giustino2007} 
relies on the spatial localization of the scattering potential and of the electron wavefunctions
when expressed in a real-space Wannier representation.

While the method of Ref.~\onlinecite{Giustino2007} was successfully applied to metals and non-polar
semiconductors \cite{Zhang2015, Louie2014PRL, Bonini2014, Noffsinger2012, Vukmirovic2012, 
Giustino2010diam, Giustino2008nature, Giustino2009graphene}, 
the same strategy breaks down in the case of polar materials. In fact the singularity
in Eq.~(\ref{M}) implies that the scattering potential is long-ranged in real space, hence the
\Frohl vertex is not amenable to Wannier-Fourier interpolation. By the same token less refined 
linear interpolation strategies are equally inadequate.

In order to deal with the polar singularity we separate the short- ($\mathcal S$) and long-range 
($\mathcal L$) contributions to the matrix element:
  \begin{equation}\label{eq.split}
  g_{mn\nu} (\mathbf k,\mathbf q) = g^{\mathcal S}_{mn\nu}(\mathbf k,\mathbf q) 
  + g^{\mathcal L}_{mn\nu}(\mathbf k,\mathbf q).
  \end{equation}
If all contributions leading to the long-wavelength divergence are collected inside
$g^{\mathcal L}$, then the short-range component will be regular and amenable to
Wannier-Fourier interpolation. This strategy is analogous to the calculation of LO-TO 
splittings in polar materials by separating the analytical and non-analytical parts of the 
dynamical matrix \cite{Gonze1997}. 

  \begin{figure*}
  \begin{center} 
  \hspace*{-10pt} \includegraphics[width=1.01\textwidth]{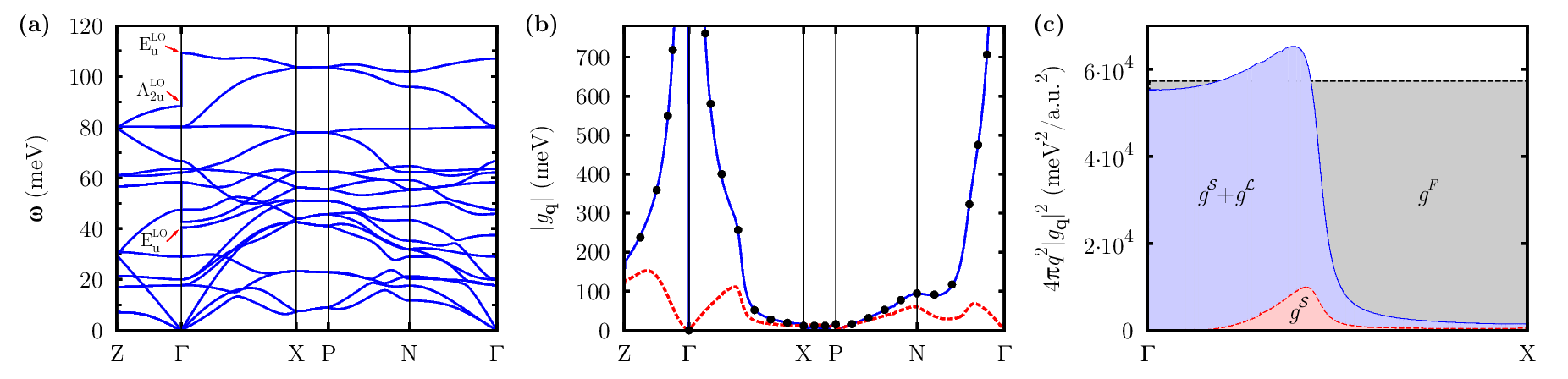}
  \caption{ \label{fig1}
  (a) Calculated phonon dispersions in anatase TiO$_2$ along high-symmetry lines in the Brillouin zone.
  The LO phonons discussed in the main text are highlighted by arrows.
  (b) Calculated electron-phonon matrix elements, with (blue solid lines) and without (red dashed lines)
  the polar coupling $g^{\mathcal L}$ from Eq.~(\protect\ref{eq.cv}). The calculations using Wannier-Fourier
  interpolations (lines) are compared to direct DFPT calculations at each wavevector (filled discs). 
  Here we show the gauge-invariant trace of $|g|^2$ over degenerate states.
  In the calculation of $g_{mn\nu}({\bf k},{\bf q})$ we set the initial electronic state $\ket{\psi_{n\mathbf k}}$
  to the bottom of the conduction band at $\Gamma$, the final electronic state $\ket{\psi_{m\mathbf k+\mathbf q}}$
  to the bottom of the conduction band, and the phonon branch to be the highest (LO) optical mode.
  (c) Spherical average of the electron-phonon matrix elements, $4\pi q^2|g|^2$, with (blue) 
  and without (red) the polar coupling, and using the simple \Frohl model in Eq.~\eqref{M} (gray). }
  \end{center} 
  \end{figure*}

\begin{figure*}
  \begin{center}
  \includegraphics[width=0.99\textwidth]{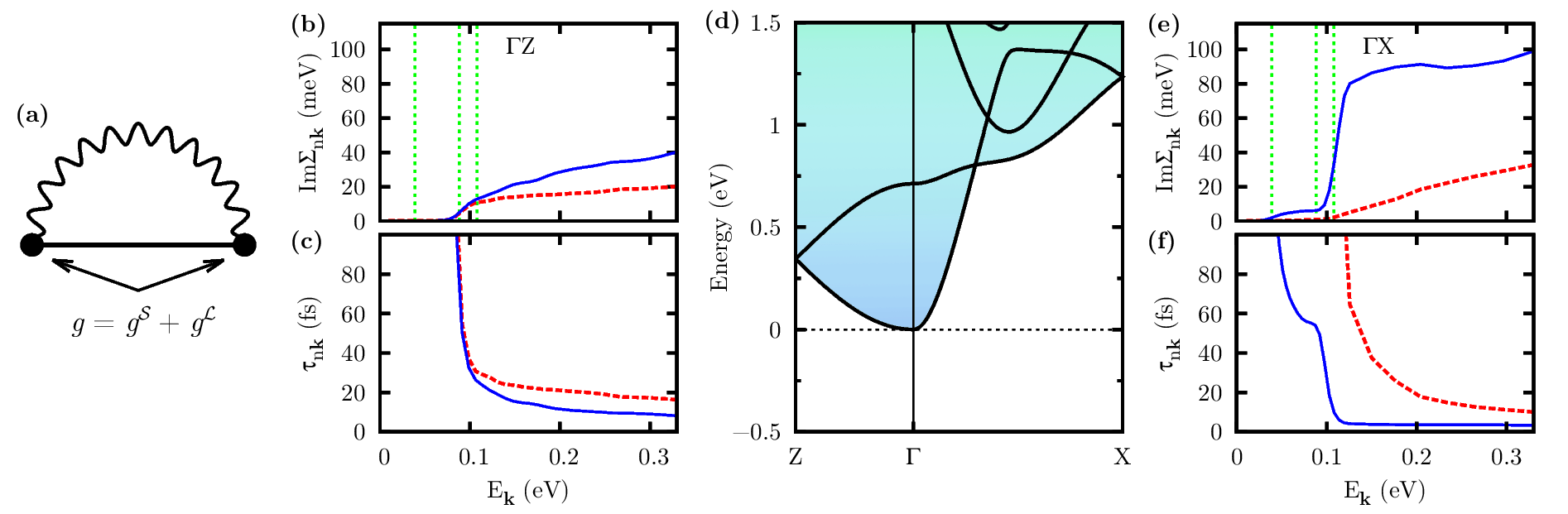}
  \caption{\label{fig2}
  (a) Diagram of the Migdal self-energy $\Sigma$ used to calculate electron lifetimes. 
  The straight and wiggly lines represent the electron and phonon Green's functions, respectively. 
  The circles are the electron-phonon matrix elements. (d) Close-up of the conduction bands of anatase 
  TiO$_2$ near the band bottom at $\Gamma$, taken as the zero of the energy. 
  (b), (e) Electron linewidths arising from electron-phonon 
  scattering, ${\rm Im}\Sigma$, along the $\Gamma Z$ and the $\Gamma X$ lines respectively, 
  near the band bottom. The energies of the LO phonons shown in Fig.~\protect\ref{fig1}(a) are 
  indicated by vertical dashed lines. (c), (f)  Electron lifetimes from (b) and (e). 
  In (b), (e) and (c), (f) the blue solid lines are computed using the complete vertex, while the 
  red dashed lines are obtained using standard Wannier interpolation \cite{Giustino2010}.  } 
  \end{center} 
  \end{figure*}

We now derive an expression for $g^{\mathcal L}$ starting from the following {\it ansatz}:
the macroscopic electric field generated by the nuclei and experienced by the electrons
can be obtained by associating an electric point dipole ${\bf p} = e\,{\bf Z}^* \cdot {\bf u}$
to each atom, where ${\bf Z}^* = Z^*_{\alpha\beta}$ is the Born effective charge tensor and 
${\bf u}=u_\alpha$ is the displacement from equilibrium [here and in the following Greek indices 
indicate Cartesian coordinates, and we use the notations $({\bf B}\cdot {\bf c})_{\alpha} = 
{\sum}_\beta B_{\alpha\beta} c_\beta$,  ${\bf a}\cdot {\bf B} \cdot {\bf c} = 
{\sum}_{\alpha\beta} a_\alpha B_{\alpha\beta} c_\beta$]. This notion draws from 
the very definition of Born charges as the sources of the macroscopic polarization \cite{BornHuang}. 
Our ansatz amounts to following similar steps 
as in the original work of \Frohl \cite{Frohlich1954}, although we are replacing ionic point 
charges by Born effective charge tensors.
A more formal theory of polar electron-phonon coupling can be developed by starting from 
a many-body approach \cite{Vogl1976}, and using the analytical 
properties of the dielectric matrix \cite{Pick1970}.
Since in polar insulators the atomic oscillations around equilibrium 
take place over timescales which are much longer
than the electronic response time, we can assume following Fr\"ohlich~\cite{Frohlich1954} that 
the electrostatic potential generated by the dipole ${\bf p}$ is screened by the high-frequency
(electronic) permittivity. This choice corresponds to assuming the adiabatic approximation. 
In the most general case of anisotropic solid this will be given by
the tensor $\bm\epsilon_\infty = \epsilon_{\infty,\alpha\beta}$. By solving the anisotropic
Poisson's equation with the dipole ${\bf p}$ placed at the origin of the reference frame we find:
  \begin{equation}\label{eq.dip1}
  V^{\mathcal L}({\bf r}) = 
    i\frac{4\pi}{N \Omega} \frac{e }{4\pi\varepsilon_0} {\bf p}\,\cdot\!
   \sum_{\bf q}\!\sum_{{\bf G}\ne -{\bf q}}\frac{ ({\bf q}+{\bf G}) \,
   e^{i({\bf q}+{\bf G})\cdot{\bf r}}}
   {({\bf q}+{\bf G})\!\cdot\!\bm\epsilon^\infty\!\cdot({\bf q}+{\bf G})},
  \end{equation}
where ${\bf G}$ indicates a reciprocal lattice vector, and the wavevectors {\bf q} belong to 
a regular grid of $N$ points in the Brillouin zone. This result is derived in the Supplemental 
Material \setcounter{footnote}{1}\footnote{See Supplemental Material at [url] for details on the 
derivation of Eq.~\eqref{eq.dip1} and Eq.~\eqref{eq.cv}, supplemental figure~1, and a 
discussion of the anisotropic linewidths in Fig.~\ref{fig2}.}. 
Now we consider that one contribution in the form of Eq.~(\ref{eq.dip1}) 
arises from each atom $\kappa$ in the position $\bm\tau_{\kappa {\bf R}} = \bm\tau_{\kappa}+{\bf R}$, 
where ${\bf R}$ denotes a lattice vector. For a given phonon with wavevector ${\bf q}$ belonging 
to the branch $\nu$ the atomic displacement pattern is given by $\Delta\bm\tau_{\kappa {\bf R}}^{({\bf q}\nu)}  
= \left(\hbar/2 N M_\kappa \omega_{{\bf q}\nu} \right)^{\!\frac{1}{2}} e^{i{\bf q}\cdot{\bf R}}\, 
{\bf e}_{\kappa\nu}({\bf q})$. In this expression ${\bf e}_{\kappa\nu}({\bf q})$ represents 
a vibrational eigenmode normalized within the unit cell. If we make the replacement ${\bf p}\rightarrow 
e {\bf Z}^*_\kappa \Delta\bm\tau_{\kappa {\bf R}}^{({\bf q}\nu)}$ inside Eq.~(\ref{eq.dip1}), 
we obtain our main result for the matrix element $g^{\mathcal L}$:
  \begin{widetext}
  \begin{equation}\label{eq.cv}
  g_{mn\nu}^{\mathcal L}({\bf k},{\bf q}) = 
  i\frac{4\pi}{\Omega} \frac{e^2 }{4\pi\varepsilon_0} \sum_{\kappa}
  \left(\frac{\hbar}{2 {N M_\kappa \omega_{{\bf q}\nu}}}\right)^{\!\!\frac{1}{2}}
  \sum_{{\bf G}\ne -{\bf q}}\frac{ ({\bf q}+{\bf G})\cdot{\bf Z}^*_\kappa \cdot {\bf e}_{\kappa\nu}({\bf q}) }
  {({\bf q}+{\bf G})\cdot\bm\epsilon^\infty\!\cdot({\bf q}+{\bf G})}
  \langle \psi_{m{\bf k+q}}| e^{i({\bf q}+{\bf G})\cdot{\bf r}} |\psi_{n{\bf k}} \rangle,
  \end{equation}
  \end{widetext}
where the bracket is to be evaluated within the Born-von K\'arm\'an supercell. Details about this 
derivation are provided in the Supplemental Material \cite{Note2}. If we consider the more restrictive 
situation of an isotropic dielectric, we find that Eq.~(\ref{eq.cv}) reduces correctly to the 
\Frohl vertex in Eq.~(\ref{M}). This result can be obtained by using the relation between the Born 
charges and the static and high-frequency permittivities \cite{Keldysh}, and by invoking the 
Lyddane-Sachs-Teller relations \cite{LST1941, *Cochran1962}. Since $g^{\mathcal L}$ reduces to the \Frohl limit 
under the assumptions used in the original work \cite{Frohlich1954}, Eq.~(\ref{eq.cv}) represents
the generalization of the \Frohl vertex for {\it ab initio} calculations. 
We note that any choice of the polar electron-phonon
coupling which did not have exactly the same limit as Eq.~(4) for ${\bf q}\to 0$ would
yield a {\it phonon} self-energy and hence phonon frequencies without the correct LO-TO splitting. 
In fact, in the long wavelength limit our ansatz leads precisely to the 
electron-phonon vertex of Ref.~\onlinecite{Vogl1976}. The advantage of our formulation is that 
the physics behind the polar singularity becomes transparent.
One interesting property of our polar vertex in Eq.~(\ref{eq.cv}) over Eq.~(\ref{M}) is that 
it naturally takes into account the periodicity of the lattice, which is not the case for the original 
\Frohl vertex. Furthermore we do not need to make any assumptions on which LO mode should 
be considered, since our formalism incorporates seamlessly the coupling to all modes, 
and the coupling strength is automatically suppressed whenever ${\bf Z}^*_\kappa \cdot 
{\bf e}_{\kappa\nu}({\bf q})$ is transverse to ${\bf q}+{\bf G}$. 

Taken together Eqs.~(\ref{eq.split}) and (\ref{eq.cv}) provide a practical recipe for calculating 
EPIs in polar materials. In the simplest approach one could perform calculations involving
only the polar coupling $g^{\mathcal L}$, in order to determine the magnitude of these effects.
In this case the phase factors 
$\langle \psi_{m{\bf k+q}} |e^{i({\bf q}+{\bf G})\cdot{\bf r}}| \psi_{n{\bf k}} \rangle$ can safely
be replaced by their ${\bf q+G}\rightarrow 0$ limit, $\delta_{mn}$, and the ensuing calculations 
of $g^{\mathcal L}$ become trivial post-processing operations. An example of this 
calculation is shown in Fig.~S2 in the Supplemental Material~\cite{Note2}.
A more refined strategy consists of computing the complete matrix elements $g = g^{\mathcal L} 
+ g^{\mathcal S}$ by exploiting Wannier-Fourier interpolation.
In this case we need to perform the following steps:
(i) evaluate the complete matrix elements $g$ on coarse Brillouin zone grids; (ii) subtract 
$g^{\mathcal L}$ so as to obtain the short-ranged part of the matrix element, $g^{\mathcal S}$; 
(iii)~apply Wannier-Fourier electron-phonon interpolation to the short-range matrix element, 
following Ref.~\onlinecite{Giustino2007}; (iv) add up the short-range part and the long-range 
part at arbitrary ${\bf k}$ and ${\bf q}$ points {\it after} interpolation. This strategy enables 
the calculation of millions of electron-phonon matrix elements for polar materials with 
{\it ab initio} accuracy, and at the computational cost of a standard calculation of phonon dispersions. 
In order to correctly capture the electronic~phases we stress that this procedure requires
the interpolation of~the~overlaps~in~Eq.~(\ref{eq.cv}).~This~is~achieved~by~using the rotation matrices $U_{nm\mathbf k}$
appearing in the definition of maximally localized Wannier functions \cite{Souza2001}:
$\ket{w_{m\mathbf R}}=\sum_{n\mathbf k}e^{-i\mathbf k\cdot\mathbf R}\,
U_{nm\mathbf k}\ket{\psi_{n\mathbf k}}$. 
Using this definition and by considering small ${\bf q+G}$ we obtain the result: 
  \begin{equation}
  \langle\psi_{m\mathbf k+\mathbf q}|e^{i({\bf q}+{\bf G})\cdot{\bf r}}|\psi_{n\mathbf k}\rangle =
  \left[ U_{\mathbf k+\mathbf q}\:U_{\mathbf k}^{\dagger} \right]_{mn}.
  \end{equation}
The matrices $U_{\mathbf k}$ are known at every point of the coarse grid from 
the calculation of maximally localized Wannier functions, and can be obtained at all other 
points via the interpolation of the electron Hamiltonian \cite{Yates2007}. 
  
In order to demonstrate our approach we consider the electron-phonon coupling in a prototypical
polar semiconductor, anatase TiO$_2$. Very recently \Frohl physics has been studied in this
material by means of angle-resolved photoelectron spectroscopy~\cite{Moser2013}.
Figure~\ref{fig1}(a) shows the phonon dispersion relations 
of anatase \TiO calculated using {\tt Quantum ESPRESSO}~\cite{QuantumEspresso}~\footnote{We performed 
ground-state density functional theory calculations using PBE \protect\cite{PBE1996} norm-conserving pseudopotentials 
including the Ti $3s$ and $3p$ semicore states, and a planewaves kinetic energy cutoff of 200~Ry. 
We sampled the Brillouin zone using a $6\times6\times6$ Monkhorst-Pack mesh. The phonon dispersions
were calculated using density functional perturbation theory (DFPT) \cite{Baroni2001}, 
by interpolating the frequencies obtained on a $4\times4\times4$ Brillouin zone grid.}. The polar 
coupling is manifest in the LO-TO splitting which can be seen around the $\Gamma$ point 
for the infrared-active $E_u$ and $A_{\rm{2u}}$ modes (the LO modes are highlighted by arrows in the plot). 
In Fig.~\ref{fig1}(b) we compare 
representative electron-phonon matrix elements calculated from DFPT with the result of our polar 
Wannier-Fourier interpolation, along high-symmetry lines in the Brillouin zone. Our interpolated 
matrix elements were obtained by performing explicit DFPT calculations on a coarse $4\times4\times4$ 
$\Gamma$-centered Brillouin zone grid. It is apparent that our method perfectly reproduces 
the polar singularities at $\Gamma$, and the behavior of the matrix element anywhere in the Brillouin zone. 
For comparison in the same figure we also show the short-range part of the matrix element $|g^{\mathcal S}|$
(red line) which is clearly non-singular near $\Gamma$. 
We stress that a standard interpolation strategy {\it without} taking into account the polar 
coupling completely fails in reproducing the correct behavior (see Fig.~S1 in the Supplemental Material \cite{Note2}). 
In order to quantify the importance of the polar divergence, in Fig.\ref{fig1}(c) we consider the quantity 
$4\pi q^2 |g|^2$ ($q=|{\bf q}|$). This quantity is ubiquitous in electron-phonon calculations since
most physical properties involve Brillouin zone integrations containing $|g_{mn\nu}({\bf k},{\bf q})|^2 
d{\bf q}$ \cite{Grimvall}. The short-range component (red) severely underestimates the complete coupling
strength, even after the singularity has been lifted by the volume element prefactor $4\pi q^2$. 
Similarly, when using the \Frohl model in Eq.~\eqref{M}, the coupling is significantly 
overestimated (gray). This demonstrates that the incorporation of the 
long-range coupling using Eq.~\eqref{eq.split} and \eqref{eq.cv} 
is essential for {\it ab initio} calculations of EPIs in polar materials.
For completeness Figs.~S3 and S4 in the Supplemental Material show that similar conclusions 
apply to two other prototypical polar compounds, GaN and LiF \cite{Note2}.

As a first example of application of our method we calculate the lifetimes of conduction electrons in 
anatase \TiO arising from the EPI. We determine the lifetimes $\tau_{n\mathbf k}$ from the imaginary part 
of  the Migdal electron self-energy $\Sigma_{n\mathbf k}$ [Fig.~\ref{fig2}(a)], using $\tau_{n\mathbf k}
=\hbar/2\,{\rm Im}\,\Sigma_{n\mathbf k}$ \cite{Mahan}. 
We consider electrons in the conduction band along the $\Gamma Z$ and the $\Gamma X$ high-symmetry 
lines, with energies near the band bottom [Fig.~\ref{fig2}(d)]; we evaluate the self-energy using 
512,000 inequivalent phonon wavevectors ($80\times80\times80$ random grid), a broadening of 10 meV, and 
a temperature of 20~K. The linewidths [Fig.~\ref{fig2}(b) and (d)] are negligible below 90~meV: this 
energy corresponds to the threshold for the emission of the high-frequency $A_{2u}$ LO 
phonon at $\Gamma$.  We can also resolve a weaker coupling with a threshold of 40~meV, associated 
with a $E_u$ LO phonon. 
Therefore, for electronic states close to the bottom of the conduction band 
we find that the coupling is dominated by the $A_{2u}$ and $E_u$ LO modes around 100~meV
(at 20~K), in agreement with the experimental findings of Ref.~\cite{Moser2013}.
We note that in this case electron-electron interactions do not contribute
to the linewidths (within the $G_0W_0$ approximation) since the electron energy
is below the energy thresholds for electron-hole pair 
generation (the fundamental gap) and for plasmon emission.
Turning to the lifetimes, we see in 
Fig.~\ref{fig2}(c) and (f) that these are essentially infinite below the phonon emission theshold, 
but they are reduced to $\sim$3--9~fs for more energetic electrons. 
Interestingly the lifetimes are highly anisotropic, varying by a factor of 
three going from $Z$ to $X$. This is related to the anisotropy of the band structure, 
and to the fact that the coupling is strongest for small phonon wavevectors~\cite{Note2}. 
For comparison we also show in Fig.~\ref{fig2} calculations performed 
with the standard Wannier interpolation technique \cite{Giustino2010}, which fails to correctly 
reproduce the matrix elements. The lifetimes are incorrectly enhanced 
by up to an order of magnitude; at the same time the anisotropy is mostly washed out [red curves 
in Fig.~\ref{fig2}(c) and (f)]. This comparison demonstrates the importance of the polar singularity 
in the correct calculation of electron lifetimes in TiO$_2$. We expect to find similar effects 
in related properties, such as polaron binding energy and carrier mobilities.

In conclusion, we introduced a method for studying electron-phonon
interactions in polar semiconductors and insulators. Our method generalizes the \Frohl theory via
a consistent description of short-range and long-range contributions to the coupling strength.
The present formalism can be employed either for complementing first-principles calculations 
at no extra cost, or for performing efficient and accurate {\it ab initio} calculations using
Wannier-Fourier interpolation. We expect that our approach will enable the calculation 
of many properties beyond the reach of current methods, including temperature dependent mobilities 
in polar semiconductors, dynamics of photoexcited carriers, and superconductivity in doped oxides.

\begin{acknowledgments}
This work was supported by the Leverhulme Trust (Grant RL-2012-001) and the UK Engineering and 
Physical Sciences Research Council (Grant No. EP/J009857/1).
This work used the ARCHER UK National Supercomputing Service via the ’AMSEC’ Leadership
project, and the Advanced Research Computing facility of the University of Oxford. \\
\textit{Note added.} --- Recently a related study of polar electron-phonon couplings 
was reported~\cite{Calandra2015}.
\end{acknowledgments}

\bibliography{biblio}

\clearpage



\section*{Supplementary material}

# Fröhlich electron-phonon vertex from first principles

*Supplemental Material*

Carla Verdi and Feliciano Giustino
*Department of Materials, University of Oxford, Parks Road, Oxford OX1 3PH, United Kingdom*

## POINT DIPOLE IN ANISOTROPIC DIELECTRIC MEDIUM

Here we derive Eq. (3) of the main text. Let us consider a point charge $Q$ placed at the position $\boldsymbol{\tau}$. We also include a uniform compensating background $-Q/N\Omega$ so that the supercell is neutral overall (the supercell consists of $N$ unit cells). The medium is an anisotropic dielectric with the permittivity tensor $\boldsymbol{\epsilon} = \epsilon_{\alpha\beta}$. The potential generated by this point charge and the background is obtained as the solution of the Poisson equation [1]:

$$(\nabla \cdot \boldsymbol{\epsilon} \cdot \nabla)\, \phi(\mathbf{r}; \boldsymbol{\tau}) = -\frac{1}{\varepsilon_0} \sum_{\mathbf{T}} \left[ Q\, \delta(\mathbf{r} - \boldsymbol{\tau} - \mathbf{T}) - \frac{Q}{N\Omega} \right], \quad (\text{S1})$$

where $\mathbf{T}$ are the lattice vectors of the Born-von Kármán supercell. The electrostatic potential $\phi$ is periodic in the supercell, therefore it can be expanded in planewaves as: $\phi(\mathbf{r}; \boldsymbol{\tau}) = \sum_{\mathbf{q},\mathbf{G}} e^{i(\mathbf{q}+\mathbf{G})\cdot\mathbf{r}} \phi(\mathbf{q}+\mathbf{G}; \boldsymbol{\tau})$, where the $\mathbf{q}$ vectors belong to a regular grid of $N$ points in the Brillouin zone, and the $\mathbf{G}$ vectors belong to the reciprocal lattice of the crystal unit cell. By replacing this expansion inside Eq. (S1) we obtain, apart from a constant:

$$\phi(\mathbf{r}; \boldsymbol{\tau}) = \frac{4\pi}{\Omega} \frac{Q}{4\pi\varepsilon_0} \frac{1}{N} \sum_{\mathbf{q}} \sum_{\mathbf{G}\neq -\mathbf{q}} \frac{e^{i(\mathbf{q}+\mathbf{G})\cdot(\mathbf{r}-\boldsymbol{\tau})}}{(\mathbf{q}+\mathbf{G})\cdot\boldsymbol{\epsilon}\cdot(\mathbf{q}+\mathbf{G})}. \quad (\text{S2})$$

The electrostatic potential generated by a dipole $\mathbf{p} = Q\mathbf{u}$ centered at the origin of the reference frame is simply given by $\phi_{\text{dip}}(\mathbf{r}) = \lim_{\mathbf{u}\to 0} \phi(\mathbf{r}; \mathbf{u}) - \phi(\mathbf{r}; 0)$:

$$\phi_{\text{dip}}(\mathbf{r}) = -i\frac{4\pi}{\Omega} \frac{1}{4\pi\varepsilon_0} \frac{1}{N} \sum_{\mathbf{q}} \sum_{\mathbf{G}\neq -\mathbf{q}} \mathbf{p} \cdot \frac{(\mathbf{q}+\mathbf{G})\, e^{i(\mathbf{q}+\mathbf{G})\cdot\mathbf{r}}}{(\mathbf{q}+\mathbf{G})\cdot\boldsymbol{\epsilon}\cdot(\mathbf{q}+\mathbf{G})}, \quad (\text{S3})$$

which is the same as Eq. (3) of the main text after setting $V^{\mathcal{L}} = -e\phi$.

## LONG-RANGE COMPONENT OF THE ELECTRON-PHONON MATRIX ELEMENT

Here we comment on the derivation of Eq. (4) in the main text. The perturbation potential entering the matrix element $g^{\mathcal{L}}$ is obtained by replacing the dipole $\mathbf{p}$ in Eq. (S3) with $e\mathbf{Z}^*_\kappa \cdot \Delta\boldsymbol{\tau}^{(\mathbf{q}\nu)}_{\kappa\mathbf{R}}$, for each atom $\kappa$ in the position $\boldsymbol{\tau}_{\kappa\mathbf{R}} = \boldsymbol{\tau}_\kappa + \mathbf{R}$. Here $\mathbf{R}$ is a lattice vector within the Born-von Kármán supercell. As stated in the main text the displacement of each atom in the vibrational eigenmode $(\mathbf{q}, \nu)$ is given by $\Delta\boldsymbol{\tau}^{(\mathbf{q}\nu)}_{\kappa\mathbf{R}} = (\hbar/2NM_\kappa\omega_{\mathbf{q}\nu})^{\frac{1}{2}} e^{i\mathbf{q}\cdot\mathbf{R}} \mathbf{e}_{\kappa\nu}(\mathbf{q})$. By combining these relations with Eq. (S3) we obtain immediately:

$$V^{\mathcal{L}}_{\mathbf{q}\nu}(\mathbf{r}) = i\frac{4\pi}{\Omega} \frac{e^2}{4\pi\varepsilon_0} \sum_\kappa \left(\frac{\hbar}{2NM_\kappa\omega_{\mathbf{q}\nu}}\right)^{\frac{1}{2}} \sum_{\mathbf{G}\neq -\mathbf{q}} \frac{(\mathbf{q}+\mathbf{G}) \cdot \mathbf{Z}^*_\kappa \cdot \mathbf{e}_{\kappa\nu}(\mathbf{q})\, e^{i(\mathbf{q}+\mathbf{G})\cdot(\mathbf{r}-\boldsymbol{\tau}_\kappa)}}{(\mathbf{q}+\mathbf{G})\cdot\boldsymbol{\epsilon}\cdot(\mathbf{q}+\mathbf{G})}. \quad (\text{S4})$$

Here we made use of the completeness relation: $\sum_{\mathbf{R}} e^{i(\mathbf{q}-\mathbf{q}')\cdot\mathbf{R}} = N\delta_{\mathbf{q}\mathbf{q}'}$, with $\delta$ the Kronecker symbol. The matrix elements is calculated as $\langle\psi_{m\mathbf{k}+\mathbf{q}}|V^{\mathcal{L}}_{\mathbf{q}\nu}(\mathbf{r})|\psi_{n\mathbf{k}}\rangle$:

$$g^{\mathcal{L}}_{mn\nu}(\mathbf{k}, \mathbf{q}) = i\frac{4\pi}{\Omega} \frac{e^2}{4\pi\varepsilon_0} \sum_\kappa \left(\frac{\hbar}{2NM_\kappa\omega_{\mathbf{q}\nu}}\right)^{\frac{1}{2}} \sum_{\mathbf{G}\neq -\mathbf{q}} \frac{(\mathbf{q}+\mathbf{G}) \cdot \mathbf{Z}^*_\kappa \cdot \mathbf{e}_{\kappa\nu}(\mathbf{q})}{(\mathbf{q}+\mathbf{G}) \cdot \boldsymbol{\epsilon} \cdot (\mathbf{q}+\mathbf{G})} \langle\psi_{m\mathbf{k}+\mathbf{q}}|e^{i(\mathbf{q}+\mathbf{G})\cdot(\mathbf{r}-\boldsymbol{\tau}_\kappa)}|\psi_{n\mathbf{k}}\rangle. \quad (\text{S5})$$

As usual the braket is to be evaluated within the Born-von Kármán supercell. Equation (4) in the main text is obtained from this expression by taking the limit $e^{-i(\mathbf{q}+\mathbf{G})\cdot\boldsymbol{\tau}_\kappa} \to 1$ inside the braket.

In principle the definition of $g^{\mathcal{L}}$ given in Eq. (S5) could be altered to include only one reciprocal lattice vector in the summation, namely $\tilde{\mathbf{G}}_{\mathbf{q}}$ such that $|\mathbf{q} + \tilde{\mathbf{G}}_{\mathbf{q}}| = \min_{\mathbf{G}} |\mathbf{q} + \mathbf{G}|$. This would correspond to retaining only the residue of the pole in reciprocal space. In practice this choice is not convenient since it leads to a Fröhlich vertex with discontinuous **q**-derivatives at the Brillouin zone boundaries. Instead, in order to minimize the number of **G** vectors in Eq. (S5) while preserving a smooth (i.e. infinitely differentiable) Fröhlich matrix element, it is sufficient to truncate the sum to the smallest shell of **G** vectors which surrounds **q**. In those cases where one is interested in **q** vectors within the first Brillouin zone, the above choice corresponds to retaining only the handful of reciprocal lattice vectors which define the first zone.

## ANISOTROPY OF THE ELECTRON LINEWIDTHS

The electron linewidths are obtained from the imaginary part of the self-energy $\Sigma$ arising from the electron-phonon coupling, calculated within the Migdal approximation [2]:

$$\Sigma_{n\mathbf{k}} = \sum_{m\nu\mathbf{q}} |g_{mn\nu}(\mathbf{k}, \mathbf{q})|^2 \left[ \frac{n_{\mathbf{q}\nu} + f_{m\mathbf{k}+\mathbf{q}}}{\epsilon_{n\mathbf{k}} - \epsilon_{m\mathbf{k}+\mathbf{q}} + \hbar\omega_{\mathbf{q}\nu} - i\eta} + \frac{n_{\mathbf{q}\nu} + 1 - f_{m\mathbf{k}+\mathbf{q}}}{\epsilon_{n\mathbf{k}} - \epsilon_{m\mathbf{k}+\mathbf{q}} - \hbar\omega_{\mathbf{q}\nu} - i\eta} \right] . \quad (S6)$$

Here $f_{m\mathbf{k}+\mathbf{q}}$ and $n_{\mathbf{q}\nu}$ are the Fermi-Dirac and the Bose-Einstein occupation numbers, respectively, $\epsilon_{n\mathbf{k}}$ and $\epsilon_{m\mathbf{k}+\mathbf{q}}$ are electron energies, $\hbar\omega_{\mathbf{q}\nu}$ is the energy of a phonon with wavevector **q** and polarization $\nu$, and $\eta$ is a small broadening (10 meV).

Figures 2(b) and (e) of the main text show the calculated linewidths of conduction electrons in anatase $TiO_2$ near the band bottom at $\Gamma$. These linewidths are highly anisotropic, varying by a factor of three when going from the $Z$ point to $X$. This feature is related to the anisotropy of the band structure, and to the fact that the coupling is strongest for small phonon wavevectors. In fact, after we take the limit of vanishing temperature in Eq. (S6), the imaginary part of the self-energy yields a factor $\delta(\epsilon_{n\mathbf{k}} - \epsilon_{m\mathbf{k}+\mathbf{q}} - \hbar\omega_{\mathbf{q}\nu})$ that signifies the energy conservation in the scattering process. In the direction $\Gamma X$, the large curvature of the band makes it possible to have electronic transitions with small momentum transfer. In the direction $\Gamma Z$ the small curvature requires electronic transitions with larger momentum transfer in order to conserve energy. As the polar matrix element goes as $1/q$, the different curvatures lead to enhanced linewidths for electrons with momentum along $\Gamma X$, as compared to electrons with momentum along $\Gamma Z$. This effect is not correctly captured by using a standard Wannier-Fourier interpolation [red lines in Fig. 2(b) and (e) in the main text]. In particular, the linewidths are systematically underestimated in the standard interpolation, and the error is larger for the linewidths along $\Gamma X$, as they involve coupling to phonons with small wavevectors.

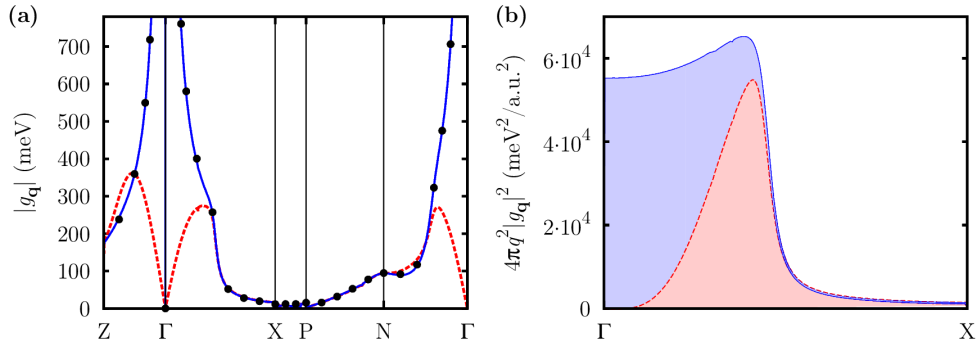

FIG. S1. (a) Calculated electron-phonon matrix elements using our polar Wannier-Fourier interpolation of Eq. (2) in the main text (blue solid lines) and the standard interpolation [3] (red dashed lines). The calculations using Wannier-Fourier interpolations (lines) are compared to direct DFPT calculations at each wavevector (filled discs) as in Fig. 1(b) in the main text. (b) Spherical average of the electron-phonon matrix elements, $4\pi q^2 |g|^2$, with the polar Wannier-Fourier interpolation (blue line) and with the standard interpolation (red line). The standard interpolation corresponds to applying the Wannier-Fourier interpolation technique of Ref. [3] to the entire electron-phonon vertex $g$ on the l.h.s. of Eq. (2). The matrix element $g$ from density-functional perturbation theory goes as $1/|\mathbf{q}|$ for $\mathbf{q} \to 0$, and vanishes at $\mathbf{q} = 0$ due to the periodic boundary conditions. As a result the Wannier-Fourier method effectively tries to interpolate a strongly fluctuating function, leading to a spurious dip at $\mathbf{q} = 0$ and unphysical behavior elsewhere in the Brillouin zone. This problem is solved by separating the vertex into long- and short-range components, and performing Wannier-Fourier interpolation only on the short-range part $g^{\mathcal{S}}$.

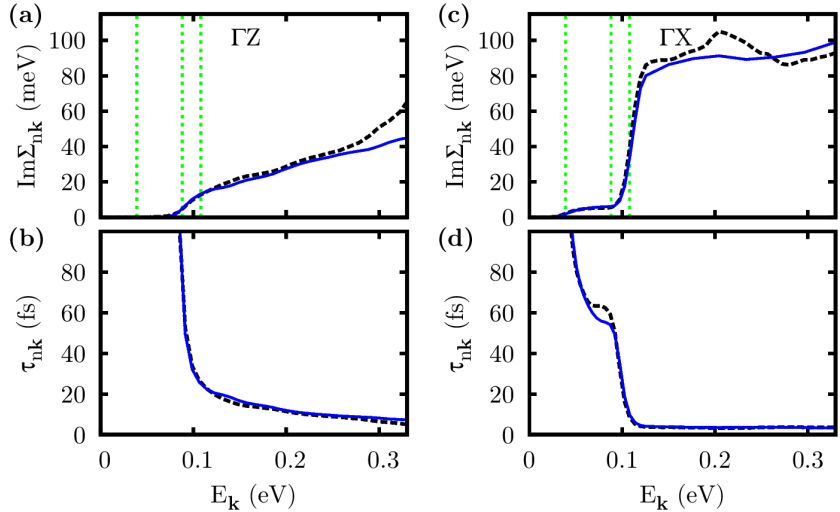

FIG. S2. (a), (c) Electron linewidths in anatase TiO$_2$ arising from electron-phonon scattering, Im$\Sigma$, along the $\Gamma Z$ and the $\Gamma X$ lines respectively, near the band bottom [as in Fig. 2(b) and (e) of the main text]. The energies of the LO phonons shown in Fig. 1(a) of the main text are indicated by vertical dashed lines. (b), (d) Electron lifetimes from (a) and (c). In (a), (c) and (b), (d) the blue solid lines are computed using the complete electron-phonon vertex $g$ in Eq. (2), while the black dashed lines are obtained by using only the long-range part of the electron-phonon vertex, $g^{\mathcal{L}}$ in Eq. (2). The two approaches yield very similar results in this case. The agreement between calculations performed using $g$ or $g^{\mathcal{L}}$ implies that the electron lifetimes in anatase TiO$_2$ are dominated by polar interactions with optical phonons.

4

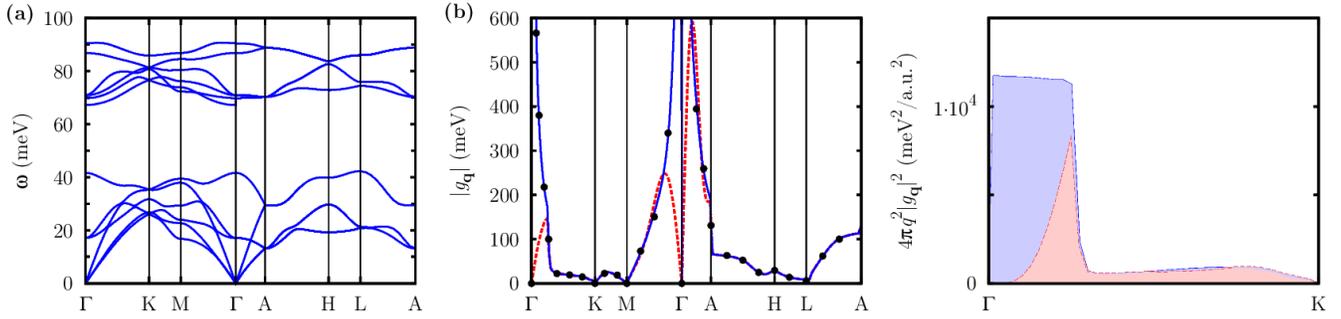

FIG. S3. (a) Calculated phonon dispersions in w-GaN along high-symmetry lines in the Brillouin zone. The phonons are interpolated from a $6 \times 6 \times 6$ Brillouin zone grid. (b) Calculated electron-phonon matrix elements using the present method (blue solid lines) and the standard interpolation [3] (red dashed lines) starting from $6 \times 6 \times 6$ electron and phonon grids. These calculations are compared to direct DFPT calculations at each wavevector (filled discs). Here we show the gauge-invariant trace of $|g|^2$ over degenerate states. In the calculation of $g_{mn\nu}(\mathbf{k}, \mathbf{q})$ we set the initial electronic state $|\psi_{n\mathbf{k}}\rangle$ to the top of the valence band at $\Gamma$, the final electronic state $|\psi_{m\mathbf{k}+\mathbf{q}}\rangle$ to the top of the valence band, and the phonon branch to be the highest (LO) optical mode. (c) Spherical average of the electron-phonon matrix elements, $4\pi q^2 |g|^2$, with the present method (blue) and with the standard interpolation (red). These calculations were performed using the lattice parameter $a = 3.158$ Å.

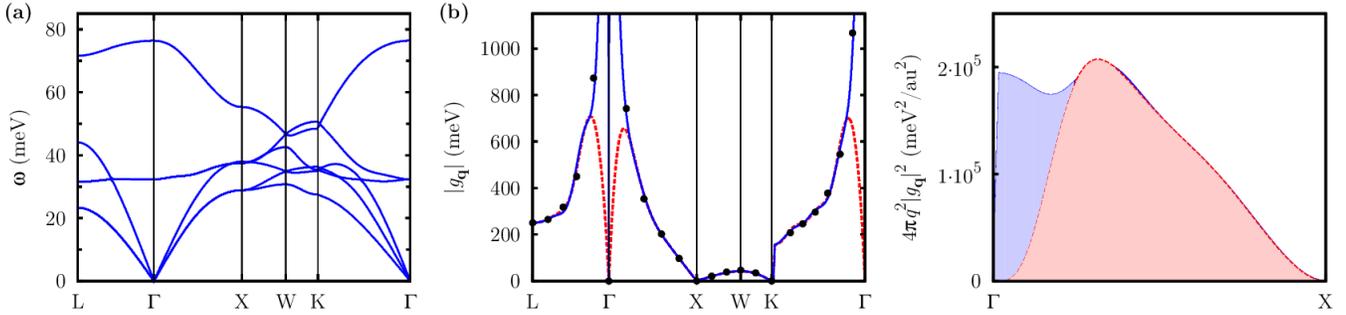

FIG. S4. (a) Calculated phonon dispersions in rocksalt LiF along high-symmetry lines in the Brillouin zone. The phonons are interpolated from a $6 \times 6 \times 6$ Brillouin zone grid. (b) Calculated electron-phonon matrix elements using the present method (blue solid lines) and the standard interpolation [3] (red dashed lines) starting from $8 \times 8 \times 8$ electron and phonon grids. These calculations are compared to direct DFPT calculations at each wavevector (filled discs). As in Fig. S3 we show the gauge-invariant trace of $|g|^2$ over degenerate states, we set the initial electronic state to the top of the valence band at $\Gamma$, the final electronic state to the top of the valence band, and the phonon branch to be the highest (LO) optical mode. (c) Spherical average of the electron-phonon matrix elements, $4\pi q^2 |g|^2$, with the present method (blue) and with the standard interpolation (red). These calculations were performed using the lattice parameter $a = 4.026$ Å.



\end{document}